# Individual and Product-Related Antecedents of Electronic Word-of-Mouth

Bogdan Anastasiei
University Alexandrul Ioan Cuza, Iasi, Romania
Faculty of Economics and Business Administration
Dept. of Management, Marketing and Business Administration
abo28@yahoo.com

Nicoleta Dospinescu
University Alexandrul Ioan Cuza, Iasi, Romania
Faculty of Economics and Business Administration
Dept. of Management, Marketing and Business Administration
dnicole@uaic.ro

Octavian Dospinescu
University Alexandrul Ioan Cuza, Iasi, Romania
Faculty of Economics and Business Administration
Dept. of Accounting, Business Informatics and Statistics
doctav@uaic.ro

**Abstract**

This research investigates the antecedents of positive and negative electronic word-of-mouth (eWOM) propensity, as well as the impact of eWOM propensity on the intention to repurchase the product. Two types of eWOM predictors were considered: product related variables and personal factors. The data were collected through an online survey conducted on a sample of 335 Romanian subjects, and the analysis method was Structural Equation Modeling. Our findings show that personal factors – social media usage behavior, marketing mavenism and need to evaluate – are the most important antecedents of the intention to write product reviews and comments online, either positive or negative. From the product related factors, only brand trust influences the propensity to provide eWOM. Furthermore, both positive and negative eWOM intentions are associated with the repurchase intention.

**Literature Review and Hypothesis Development**

In the online business environment, consumers often rely on reviews and ratings to make purchasing decisions (Tran, 2020), (Zhu, Li, Wang, He, & Tian, 2020). In this context, the perceived quality of a product or service plays a key role in shaping the intention to issue a positive E-WOM (Electronic Word of Mouth). Perceived quality is the subjective assessment that a consumer makes about the quality of a product or service based on their perceptions and experiences. It is a key factor in the purchasing decision making process and influences consumer attitudes and behaviour. Recent studies (Jain, Dixit, & Shukla, 2023) highlight the importance of perceived quality in determining how consumers share their experiences online through E-WOM.

According to (Madi, Al Khasawneh, & Dandis, 2024), perceived quality has a significant influence on the intention to issue positive E-WOM. When consumers perceive a product or service to be of high quality, they are more likely to share this positive experience with other potential buyers. For example, a consumer who purchases a product and experiences superior quality or outstanding performance *may be motivated to share this positive experience through online reviews or social media platforms.* Positive reviews generated by positive E-WOM intent have a significant impact on overall brand perception. Studies (Kim, Park, & Mariani, 2023) show that consumers pay particular attention to online reviews and ratings in their decision-making process.

Other recent studies (Ahmad, Abuhashesh, Obeidat, & AlKhatib, 2020) highlight that 'perceived quality' plays a significant role *in the formation of negative E-WOM intention*. When consumers perceive a decline in quality in a product or service, they are more likely to share this negative experience via online reviews or other social media platforms. For example, according to (Boo & Kim, 2013), a customer experiencing a perceived low quality of a product may be motivated to share this experience through negative comments on review sites or social media. Negative reviews and negative E-WOM can have a significant impact on a brand's reputation. Research conducted by (Yang & Ha, 2023) indicates that consumers pay particular attention to negative reviews and ratings, and these can affect the overall perception of the brand. Negative E-WOM (Roy, Datta, Mukherjee, & Basu, 2021) can raise concerns about the quality of products or services offered, which can lead to lower consumer confidence and thus lower their willingness to make further purchases from that brand.

Based on the results previously highlighted in the literature, we formulate the following research hypotheses to be included in the proposed model.

**H1: Perceived quality has a positive influence on positive e-wom intention**

**H2: Perceived quality has a negative influence on negative e-wom intention**

Brand trust is a subjective assessment that consumers make about the credibility and reliability of a brand. It is a fundamental pillar in the decision-making process and contributes to consumer loyalty and engagement. According to (Jung, Kim, & Kim, 2014), in today's digital environment, trust in a brand is key to building a lasting relationship with consumers. This trust can influence not only purchase decisions, but also the intention to issue positive E-WOM. Recent studies (Jain, Dixit, & Shukla, 2023) highlight *the positive impact of brand trust on positive E-WOM intention*. When consumers trust a brand, they are more likely to share their positive experiences with other online users. For example (Sari, Fauzi, & Rini, 2021), a consumer who has had consistently positive experiences with a brand may be motivated to express their enthusiasm through positive reviews, ratings or social media posts. According to (Seifert & Kwon, 2020), online communities formed around a brand where positive experiences are shared become natural ambassadors, helping to extend and strengthen the brand's positive reputation.

Other recent studies (Bhandari & Rodgers, What does the brand say? Effects of brand feedback to negative eWOM on brand trust and purchase intentions, 2020) indicate the importance of brand trust in the context of E-WOM, with a particular focus *on the intention to give negative feedback*. Brand trust (Bigné, Ruiz-Mafé, & Badenes-Rocha, 2023) has a significant impact on the intention to express negative E-WOM. Consumers who trust a brand may be more reluctant to express negative feedback, even when encountering negative experiences. This trust acts as a barrier to expressing dissatisfaction online. For example, a consumer who is loyal to a brand may be less likely to share a negative experience about that brand's products or services for fear of damaging their relationship with the brand. On the other hand, negative feedback may affect trust in the brand. Research conducted by (Bhandari, Rodgers, & Pan, Brand feedback to negative eWOM messages: Effects of stability and controllability of problem causes on brand attitudes and purchase intentions, 2021) indicates that negative reviews and E-WOM can erode brand trust, causing consumers to reconsider their brand loyalty. Negative reviews (Chang, Rhodes, & Lok, 2013) can raise concerns about the quality or reliability of the brand, thus affecting overall brand perception.

The results of previous research in the literature lead us to generate the following research hypotheses.

**H3: Brand trust has a positive influence on positive e-wom intention**

**H4: Brand trust has a negative influence on negative e-wom intention**

In a competitive business environment, consumers' perception of the price of a product or service (Abdullah, Febrian, Perkasa, Wuryandari, & Pangaribuan, 2023) can influence not only their purchasing decisions, but also how they share that experience online through E-WOM. Price perception reflects consumers' subjective assessment of the cost of a product or service in relation to its perceived benefits. It is a key factor in shaping perceptions of value and can strongly influence purchasing behaviour. Recent studies (Bushara, et al., 2023) highlight *the impact of price perception on the willingness to share positive experiences online through E-WOM*. Price perception can have a significant impact on the intention to issue positive E-WOM. When consumers perceive that a product or service offers excellent value for the price paid, they are more likely to share this positive experience with other consumers online. For example, a consumer who discovers a bargain offer or benefits from perceived substantial discounts may be motivated to express their enthusiasm through positive reviews or recommendations on social platforms. Studies conducted by (Sugiran, Syuhada, Sulaiman, Mas'od, & Hasbullah, 2022) and (Bai, Marsden, Ross Jr, & Wang, 2017) have shown that promotions and discounts influence price perception and thus the intention to issue positive E-WOM. Consumers are more likely to share their positive experiences when they perceive that they have benefited from a special offer or a great price. According to (Blom, Lange, & Hess, 2021), promotions can create a sense of excitement and satisfaction among consumers, stimulating them to share this positive experience online.

On the other hand, recent studies (Bambauer-Sachse & Young, 2023) highlight *the importance of price perception in the context of forming the intention to issue negative E-WOM*. Price perception can have a significant impact on the intention to express negative E-WOM. When consumers perceive that the price of a product or service is not justified relative to the benefits offered, they are more likely to share this negative experience with other online consumers. For example (Xu, Zheng, & Yang, 2023), a consumer who feels they have paid too much for a product or service and have not received the expected value may be motivated to express their dissatisfaction through negative reviews or comments on social platforms. Studies (Schneider & Huber, 2021) show that negative experiences, such as poor service or faulty products, can influence price perception and therefore the intention to issue negative E-WOM. Consumers who have had negative experiences tend to perceive prices as higher and may be more willing to share this perception with others (Golmohammadi, Mattila, & Gauri, 2020).

Based on previous relevant findings from the knowledge domain, we integrate the following research hypotheses into the model proposed in this article.

**H5: Price perception has a positive influence on positive e-wom intention**

**H6: Price perception has a negative influence on negative e-wom intention**

With the expanding use of online social media platforms, it is crucial to understand how these media influence the way consumers share their experiences. With a significant increase in the use of social media, its influence on consumer behaviour is becoming increasingly evident.

Recent studies conducted by (Leong, Loi, & Woon, 2022) and (Pang, 2021) have found *a significant association between active social media use and intention to issue positive e-WOM*. Active users of social platforms, through reviews, ratings and recommendations, contribute to the creation of an online community promoting products and services (Kilipiri, Papaioannou, & Kotzaivazoglou, 2023). This positive interaction increases the influence of social media on other users' purchasing decisions. A concrete example (Lee & Kim, 2020) is the intensive activity of influencers on Instagram, who, through genuine posts and recommendations, generate an increased intention to emit positive e-WOM among their followers.

On the other hand, recent research by (Izogo, Jayawardhena, & Karjaluoto, 2023) suggests that *social media use may also influence the intention to express negative e-WOM*. According to (Ribeiro & Kalro, 2023), disappointed or dissatisfied consumers have ample platforms to voice their grievances and warn other consumers about their negative experiences. For example (Lutzky, 2024), a dissatisfied customer may use Twitter or online forums to express their frustration with a product or service, generating negative e-WOM intent that can damage a company's reputation.

Given previous findings in the literature showing that heavy use of social media is closely related to e-WOM intention, whether positive or negative, we formulate the following research hypotheses:

**H7: Social media usage level has a positive influence on positive e-wom intention**

**H8: Social media usage has a negative influence on negative e-wom intention**

In an era where consumers increasingly rely on digital information and online interactions, the concept of marketing mavenism (Abbas, Khwaja, Abbasi, & Hameed, 2023), (Robinson Jr, 2023) has evolved to reflect the powerful influence that individuals with expertise in a particular field can have. According to (Awais, Samin, Gulzar, Hwang, & Zubair, 2020) and (Farragher, Wang, & Worsley, 2016), these mavens, be they experts in tech, fashion or gastronomy, become digital opinion leaders with a significant impact on purchasing decisions. Marketing mavenism refers to the influence that people with deep expertise in a particular field have on the buying behaviour of other consumers. In the digital age (Albinali, Han, Wang, Gao, & Li, 2016), mavens are not only recognised experts, but also online influencers or content creators who share their passion and knowledge with broad audiences. These digital mavens often become trusted sources for consumers, influencing purchasing decisions through online reviews, tutorials and recommendations. Subject matter expertise, authenticity and the ability to communicate effectively with audiences are key characteristics of mavens in the digital age. Recent research (Lee & Fiore, 2024) shows that digital marketers gain consumer trust through authentic content and relevance of information. In addition, their ability to generate conversations and actively interact with the online community helps to reinforce their status as thought leaders.

Recent literature (Goldring, Gong, & Gironda, Mavens at work: Brand commitment and the moderating role of market mavens on social media engagement, 2022) indicates that the influence of digital drivers has a significant impact on consumer behaviour. Studies (Goldring & Azab, New rules of social media shopping: Personality differences of US Gen Z versus Gen X market mavens, 2021) show that their recommendations can affect the attitudes, preferences and purchase intentions of their audiences. Digital mavens have the ability to create trends and influence purchasing decisions in a rapid and widespread way, thanks to their amplification on online social media platforms. Companies (Khan, Fazili, & Bashir, 2022) have begun to integrate mavenism marketing strategies into their promotional efforts. Collaborating with relevant industry mavens to promote products and services is becoming an essential component of marketing campaigns. Companies are turning their attention to identifying mavens who have a genuine connection to their brand values and target audience.

Recent studies (Horng, Liu, Chou, Yu, & Hu, 2024) have found that *marketing mavenism has a significant impact on the intention to issue positive e-WOM*. Mavens, through their online experience and reputation, become credible sources of information for other consumers. Positive recommendations from mavens (Quach & Lee, 2021) can influence the purchasing decisions of other users, building trust in brands and products. A notable example is the active presence of mavens in product or service reviews on platforms such as Amazon or Yelp, where the positive impact of mavens is evident in increased sales and brand awareness.

On the other hand, recent literature (Taketani & Mineo, 2021) suggests *that mavens can also influence negative e-WOM*. Even though mavens are traditionally seen as promoters of products and services, they can also share negative experiences. Criticism from mavens (Colmekcioglu, Marvi, Foroudi, & Okumus, 2022) can have a significant impact on the reputation of a business and can raise serious concerns for consumers. For example, a technology maven reporting significant shortcomings of an electronic product may cause potential buyers to avoid purchasing it, negatively impacting sales and brand image.

Based on previous findings in the literature, we propose the following research hypotheses.

**H9: Marketing mavenism has a positive influence on positive e-wom intention**

**H10: Marketing mavenism has a negative influence on negative e-wom intention**

"Need to evaluate" is the level of interest and importance attributed by consumers to the evaluation of products or services. In the digital age, where access to information is at everyone's fingertips, "need to evaluate" plays an extremely important role in consumer decision-making (Martínez-Navalón, Gelashvili, & Gómez-Ortega, 2021).

Recent studies (Amed, Mukherjee, Das, & Datta, 2019) indicate that consumers with a high level of "need to evaluate" *are more likely to express positive e-WOM*. These consumers are more detail-oriented and actively seek information about the products or services they use. Positive reviews and recommendations thus become a natural way to share their experiences with other potential consumers. A concrete example (Wen, Hu, & Kim, 2018) is the active presence on online forums and review platforms of consumers who want to share their positive experiences with products they have purchased.

On the other hand, other research findings (Lis & Fischer, 2020) suggest that *"need to evaluate" may also contribute to negative e-WOM expression*. Consumers with an increased level of "need to evaluate" may become more vocal and critical in expressing their dissatisfaction. Actively seeking details and carefully evaluating negative experiences encourages them to express their disappointments publicly to warn other consumers. For example (Xun & Guo, 2017), consumers who have experienced significant problems with products or services may choose to express their dissatisfaction on platforms such as Twitter or review sites.

Given that "need to evaluate" is a significant factor in determining the intention to issue e-WOM, whether positive reviews or negative remarks, we formulate the following research hypotheses.

**H11: Need to evaluate has a positive influence on positive e-wom intention**

**H12: Need to evaluate has a negative influence on negative e-wom intention**

In today's digital age, e-WOM has become an extremely valuable source of information for consumers. Expressing reviews and opinions online can influence purchasing decisions and have a significant impact on the success of a business.

Recent studies conducted by (Wandoko & Panggati, 2022) and (Zeqiri, Ramadani, & Aloulou, 2023) have shown *that there is a strong correlation between positive e-WOM and intention to repurchase*. Consumers who express positive reviews about a product or service are often more likely to return and make subsequent purchases from the same brand or business. According to (Bulut & Karabulut, 2018), this link is reinforced by the increased trust consumers develop in brands that receive consistent positive feedback. For example, a customer who gives a rave review for a positive online shopping experience is more likely to return and make further purchases from the same online store.

On the other hand, research by (Ginting, Chandra, Miran, & Yusriadi, 2023) has shown that *negative e-WOM intention can have a significant impact on re-purchase intention*. When consumers are exposed to negative reviews or feedback about a product or service, they are more likely to reconsider their decision to make another purchase from the same source (Sa'ait, Kanyan, & Nazrin, 2016). This may be driven by concerns about product quality or previous negative experience. For example (Johnson Jorgensen & Sorensen, 2021), a customer who has had a disappointing experience with a product or service may be discouraged from making another purchase from the same company or returning to the same brand in the future.

Given that previous findings in the literature show that e-WOM intention, whether positive or negative, has a significant impact on consumers' intention to repurchase, we formulate the following research hypotheses.

**H13: Positive e-wom intention has a positive influence on intention to re-purchase**

**H14: Negative e-wom intention has a negative influence on intention to re-purchase**

Based on the previous developed research hypotheses, the research framework is presented in Figure 1.

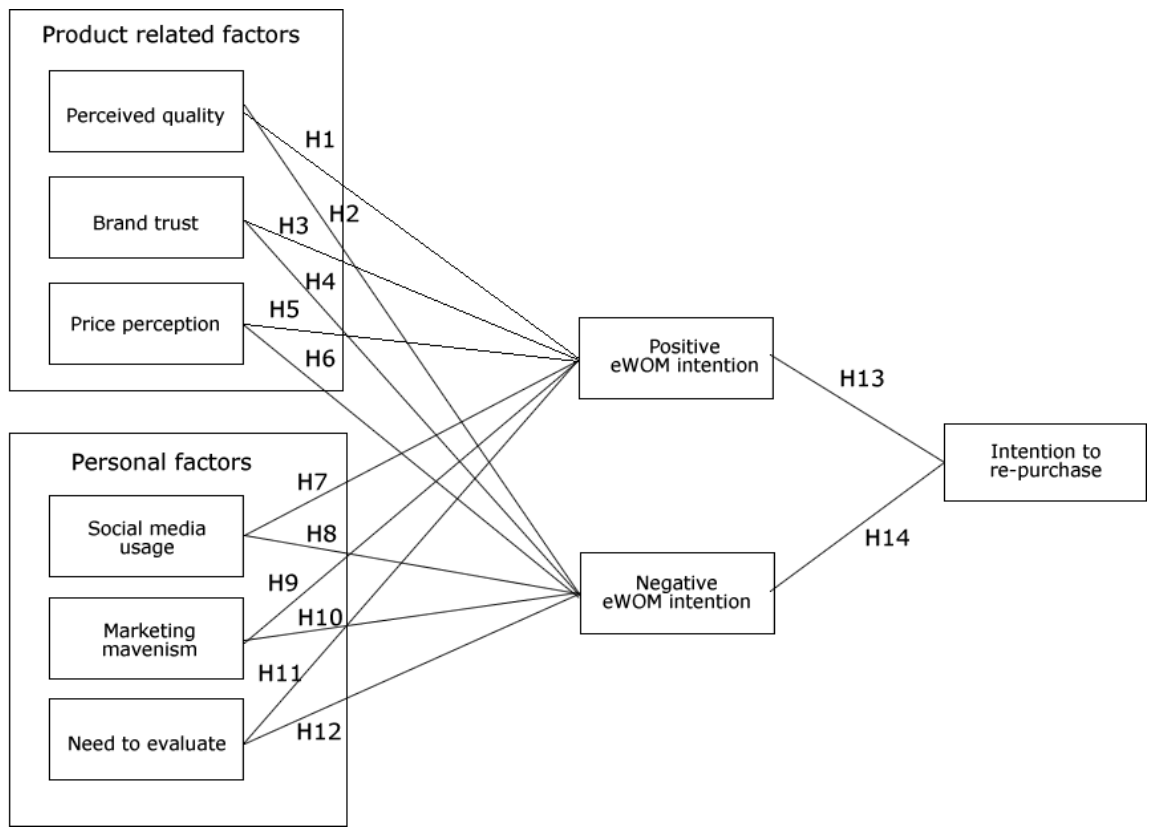

Figure 1. The research framework

**Instrument and methodology**

In order to collect our data, we conducted an online survey to a convenience sample of 335 Romanian students and young professionals. The data was collected in the period January-February 2024. The demographic structure of our sample can be seen in Table 1.

Table 1. **Sample structure by gender and age**

| Gender | |
|---|---|
| Male | 34.3% |
| Female | 65.7% |
| **Age group** | |
| 18-25 | 94.6% |
| 26-35 | 2.4% |
| 36-45 | 2.7% |
| 46-55 | 0.3% |

The respondents were invited to answer fifty-one questions, divided into nine scales. Each scale was used to measure a specific construct. The constructs used in our research model are: product quality, price perception, brand trust, social media usage, marketing mavenism, need to evaluate, intention to provide positive word-of-mouth, intention to provide negative word-of-mouth and intention to re-purchase.

To evaluate product quality, the scales of (Wu, Lu, Wu, & Fu, 2012) was used. To measure price perception and brand trust we have employed the scales proposed by (Dodds, Monroe, & Grewal, 1991) and (Chaudhuri & Holbrook, 2001). The social media usage habits were measured with the scale created by (Rapp, Beitelspacher, Grewal, & Hughes, 2013), while marketing mavenism was assessed with the scale proposed by (Steenkamp & Gielens, 2003). The need to evaluate construct was evaluated using the scale of (Jarvis & Petty, 1996).

To assess the respondents' propensity to provide positive or negative WOM in the online environment we have adapted the scales created by (Alexandrov, Lilly, & Babakus, 2013) and (Wolter & Cronin, 2016), respectively. Finally, to measure the intention to re-purchase we used the scale of (Dodds, Monroe, & Grewal, 1991). The scales used in our research are presented in Appendix 1.

**Data Analysis and Results**

Our data analysis process was divided into three big steps. In the first step we conducted an exploratory factor analysis (EFA), to find out whether the individual items are well correlated with the associated latent variables. This analysis was done using the IBM SPSS software, version 26. At the end, twenty-two items out of fifty-one were excluded, because they presented either big cross-loadings or poor loadings. More precisely, we have eliminated two price perception items, six social media usage items, one marketing mavenism item, twelve need to evaluate items and one positive e-WOM propensity item. The Kaiser-Meyer-Olkin indicator for the last EFA model was 0.890, indicating very good factor adequacy. The Bartlett's sphericity test was statistically significant ($p<0.01$).

In the second step we have executed a confirmatory factor analysis (CFA), to evaluate the relationships between our constructs and their related items. The cutoff values that we have used to assess the goodness-of-fit of the measurement model were the following: for the comparative fit index (CFI) – 0.900 (Hair, Black, Babin, & Anderson, 2010), for the Tuckey-Lewis index (TLI) – 0.900 (Hu & Bentler, 2009), for the goodness-of-fit index (GFI) – 0.800 (Greenspoon & Saklofske, 1998), for the adjusted goodness-of-fit index (AGFI) – 0.800 (Ellison, Steinfield, & Lampe, 2007), for the root mean square error of approximation (RMSEA) – 0.08 (Hair, Black, Babin, & Anderson, 2010), for the standardized root mean square residual (SRMR) – 0.08 (Hu & Bentler, 2009), for the $\chi^2/df$ ratio – between 1 and 5 (Marsh & Hocevar, 1985).

The values for our measurement model are: CFI = 0.961, TLI = 0.954, GFI = 0.881, AGFI = 0.849, RMSEA = 0.050, SRMR = 0.041, $\chi^2/df$ = 1.837. All these values meet the cutoff, so our model is very good fit.

The main indicators of the measurement model are presented in Table 2. The regression weights of the individual items are statistically significant ($t > 1.96$) and their standardized values are higher than 0.5. The average variance extracted (AVE) are also greater than 0.5, showing good convergent validity. Moreover, all latent variables have good internal consistency (Cronbach's alpha values and composite reliabilities are higher than 0.7).

Table 2. **Summary indicators of the measurement model**

| Constructs and items | Beta | t-value | SE | Alpha | Composite Reliability | AVE |
|---|---|---|---|---|---|---|
| **Product quality** | - | - | - | 0.914 | 0.826 | 0.567 |
| This product is of good quality | 0.906 | - | - | - | - | - |
| This product is of high workmanship | 0.909 | 25.764 | 0.039 | - | - | - |
| This product is durable | 0.830 | 21.142 | 0.047 | - | - | - |
| This product is reliable | 0.776 | 18.565 | 0.050 | - | - | - |
| This product is dependable | 0.701 | 15.620 | 0.059 | - | - | - |
| **Price perception** | - | - | - | 0.912 | 0.817 | 0.555 |
| At the price shown, the product is very economical | 0.899 | - | - | - | - | - |
| The product is considered a good buy | 0.902 | 23.286 | 0.042 | - | - | - |
| The price shown for the product is very acceptable | 0.843 | 20.817 | 0.048 | | | |
| **Brand trust** | - | - | - | 0.927 | 0.865 | 0.679 |
| I trust this brand | 0.914 | - | - | - | - | - |
| I rely on this brand | 0.868 | 23.616 | 0.046 | - | - | - |
| This is an honest brand | 0.834 | 21.632 | 0.045 | - | - | - |
| This brand is safe | 0.881 | 24.414 | 0.040 | - | - | - |
| **Social media usage** | - | - | - | 0.888 | 0.709 | 0.551 |
| I use social media to monitor products and brands | 0.887 | - | - | - | - | - |
| I use social media to follow promotions | 0.901 | 15.442 | 0.067 | - | - | - |
| **Marketing mavenism** | - | - | - | 0.852 | 0.711 | 0.542 |
| I like introducing new brands and products to my friends | 0.688 | - | | | | |
| My friends and neighbors often come to me for advice | 0.906 | 14.314 | 0.083 | - | - | - |
| People often ask me for my opinion about new products | 0.868 | 14.065 | 0.087 | - | - | - |
| **Need to evaluate** | - | - | - | 0.873 | 0.701 | 0.567 |
| It bothers me to remain neutral | 0.676 | - | - | - | - | - |
| I like to have strong opinions when I am not personally involved | 0.843 | 13.372 | 0.094 | - | - | - |
| I have many more opinions than the average person | 0.802 | 12.858 | 0.095 | - | - | - |
| I would rather have a strong opinion than no opinion at all | 0.868 | 13.645 | 0.093 | - | - | - |
| **Positive eWOM intention** | - | - | - | 0.920 | 0.722 | 0.599 |
| I would post positive things about the brand | 0.925 | - | - | - | - | - |
| I would recommend this brand to the people in my social network | 0.922 | 22.599 | 0.043 | - | - | - |
| **Negative eWOM intention** | - | - | - | 0.921 | 0.752 | 0.528 |
| I would complain to the members of my social network | 0.932 | - | - | - | - | - |
| I would discuss with the members of my social network about my frustrations | 0.887 | 25.075 | 0.037 | - | - | - |
| I would say negative things about the brand in my social networks | 0.861 | 23.541 | 0.037 | - | - | - |
| **Intention to re-purchase** | - | - | - | 0.879 | 0.742 | 0.530 |
| The likelihood to re-purchase this product is… | 0.885 | - | - | - | - | - |
| The probability that I would buy this product again is… | 0.799 | 17.171 | 0.046 | - | - | - |
| My willingness to buy this product again is... | 0.844 | 18.329 | 0.051 | - | - | - |

Further, to evaluate the discriminant validity of our measurement model we have compared the construct squared correlations with the average variance extracted. As it can be noticed in Table 3, all AVE values (in the main diagonal) are greater than the corresponding squared correlations, denoting good discriminant validity.

Table 3. **Average variance extracted and squared correlations**

|     | PQ    | PP    | BT    | SMU   | MM    | NE    | PI    | NI    | IRP   |
|-----|-------|-------|-------|-------|-------|-------|-------|-------|-------|
| PQ  | **0.567** |       |       |       |       |       |       |       |       |
| PP  | 0.319 | **0.555** |       |       |       |       |       |       |       |
| BT  | 0.461 | 0.272 | **0.679** |       |       |       |       |       |       |
| SMU | 0.024 | 0.081 | 0.093 | **0.551** |       |       |       |       |       |
| MM  | 0.062 | 0.026 | 0.107 | 0.250 | **0.542** |       |       |       |       |
| NE  | 0.028 | 0.040 | 0.085 | 0.090 | 0.281 | **0.567** |       |       |       |
| PI  | 0.012 | 0.032 | 0.024 | 0.210 | 0.186 | 0.172 | **0.599** |       |       |
| NI  | 0.006 | 0.002 | 0.003 | 0.113 | 0.099 | 0.140 | 0.452 | **0.528** |       |
| IRP | 0.171 | 0.113 | 0.198 | 0.088 | 0.124 | 0.073 | 0.090 | 0.009 | **0.530** |

PQ – product quality, PP – price perception, BT – brand trust, SMU – social media usage, MM – marketing mavenism, NE – need to evaluate, PI – positive eWOM intention, NI – negative eWOM intention, IRP – intention to re-purchase

During the third step of our analysis we tested our causal model, presented in Figure 1. The values of the goodness-of-fit indicators for this model were: CFI = 0.953, TLI = 0.945, GFI = 0.872, AGFI = 0.839, RMSEA = 0.054, SRMR = 0.074, $\chi^2/df$ = 1.991. These values denote very good model fit.

The path coefficients for the causal model can be inspected in Table 4 and in Figure 2.

Table 4. **Path coefficients of the causal model**

| Hypothesis | Path | Coefficient | p | Result |
| --- | --- | --- | --- | --- |
| H1 | Perceived quality -> Positive eWOM intention | 0.040 | 0.753 | Not supported |
| H2 | Perceived quality -> Negative eWOM intention | -0.100 | 0.428 | Not supported |
| H3 | Brand trust -> Positive eWOM intention | -0.162 | 0.196 | Not supported |
| H4 | Brand trust -> Negative eWOM intention | -0.401 | 0.001 | Supported |
| H5 | Price perception -> Positive eWOM intention | 0.089 | 0.373 | Not supported |
| H6 | Price perception -> Negative eWOM intention | 0.097 | 0.330 | Not supported |
| H7 | Social media usage -> Positive eWOM intention | 0.347 | <0.001 | Supported |
| H8 | Social media usage -> Negative eWOM intention | 0.270 | <0.001 | Supported |
| H9 | Marketing mavenism -> Positive eWOM intention | 0.247 | 0.026 | Supported |
| H10 | Marketing mavenism -> Negative eWOM intention | 0.157 | 0.154 | Not supported |
| H11 | Need to evaluate -> Positive eWOM intention | 0.425 | <0.001 | Supported |
| H12 | Need to evaluate -> Negative eWOM intention | 0.504 | <0.001 | Supported |
| H13 | Positive eWOM intention -> Intention to repurchase | 0.362 | <0.001 | Supported |
| H14 | Negative eWOM intention -> Intention to repurchase | -0.182 | 0.005 | Supported |

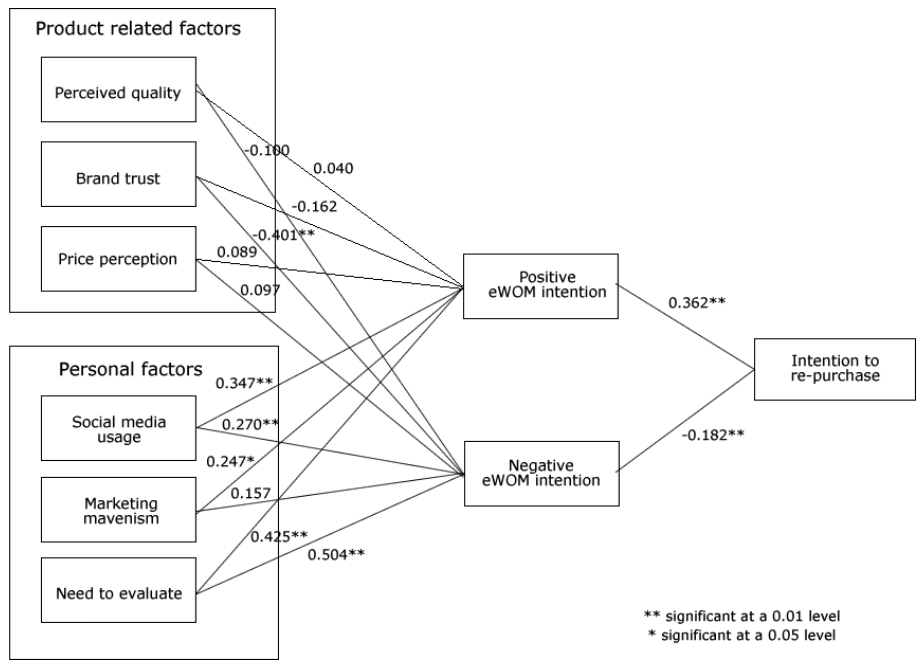

Figure 2. Path coefficients of the causal model

In conclusion, eight hypotheses out of fourteen are supported; the other six hypotheses cannot be validated. These findings will be discussed in detail in the following section.

**Discussion**

The first important observation we can make is that from all product related factors considered in our research, only brand trust influences eWOM intention. More exactly, there is a negative association between brand trust and the propensity to deliver bad eWOM online. In other words, customers with high levels of brand respect and confidence will not rant about that brand in the social media, even if they happen to be disappointed with their purchase at some point in time. Thus, brand trust acts like an inhibitor for the negative eWOM intentions. On the other hand, this variable does not affect the positive eWOM propensity. In consequence, the simple fact that customers have a good relationship with the brand does not necessarily convert them into brand evangelists on the social media platforms.

The other product related factors have no relationship with the intention to generate eWOM, either positive or negative. People who appreciate the product quality and consider it as being good value for money will not feel compelled to praise the product in the social networks and recommend it to other users. Furthermore, customers who are unhappy with quality and price will not automatically complain in the social media and warn other people about the product. In conclusion, product quality and price attractiveness do not seem to be strong enough eWOM triggers.

There is some disagreement between the above results and previous research. Many authors showed that quality and customer satisfaction are positively associated with eWOM propensity; for example (Wang, Tran, & Tran, 2017) or (Lee Y. , 2016). We deem that this divergence could be because previous studies do not consider any personality traits (or other variables that do not depend on the customer experience with the product) that might actually motivate people to write reviews and comments in the social media. Anyway, some more research may still need to be conducted before drawing firm conclusions about these hypotheses.

The personal factors have more significant influence, overall, on eWOM intentions. The social media usage behavior is positively associated with both good and bad eWOM. This result is consistent with the ones obtained by (Mikalef, Pateli, & Giannakos, 2013), who show that intense social media users are more inclined to generate word-of-mouth. Moreover, (Durukan & Bozaci, 2012) show that social media using habits are associated with both positive and negative eWOM intentions. People with high levels of social media usage employ social networks to monitor brand related events, to follow promotions, to communicate with retailers. They keep contact and have a constant dialogue with both companies and customers (their social media peers). Hence, it is natural for them to discuss product and brand related issues with other network users. As a consequence, there is a high probability that they will provide positive or negative eWOM related to their purchases. They will likely eulogize the product if they are happy, or criticize it if they are discontented.

Need to evaluate is a strong eWOM predictor as well. This personality trait reflects an individual's propensity to engage in evaluative processes and maintain strong opinions and attitudes. Our study shows that the intention to spread eWOM, either positive or negative, is positively associated with this attribute. The big need to evaluate score amounts will cause social media users to express their judgments and feelings about the products they purchased, because these people tend to form clear-cut opinions about everything and don't like to remain neutral. At the same time, people with low need to evaluate levels are pretty much indifferent with most issues, even those that affect them directly. For this reason, they will not feel the urge to discuss their frustrations with products and brands in the online networks, or to promote good products to their online peers. So these users will likely generate low quantities of eWOM. These findings are in consonance with the results of (Xu, Reczek, & Petty, 2023), who demonstrate that need to evaluate is a significant predictor of creating online word-of-mouth.

Furthermore, our results indicate that marketing mavenism only determines the propensity to provide positive eWOM. Marketing mavens are individuals that like to diffuse information in the marketplace. They often introduce new products and brands to their friends, talk about their acquisitions and offer advice to other people. Mavens don't use to talk negatively about brands; they typically recommend and endorse good products. Therefore, mavenism is associated with the intention to provide positive eWOM in the social media. These results are confirmed by the research of (Goldsmith & Clark, 2006). These authors studied the relationships between mavenism and the motives to provide online advice about products and services. Their findings show great correlations between mavenism and motives like concern for other consumers, positive self enhancement, social benefits, and desire to help the company. So more often than not, marketing mavens offer positive advice about high-quality products. At the same time, the above mentioned study found no correlation between mavenism and need to vent negative feelings. In consequence, mavens rarely use the social media platforms to rant about products or brands.

At this point we can conclude that individuals' personal traits are more important predictors for eWOM propensity than product related factors. The most probable eWOM providers are people who present at least one of these characteristics: are heavy social media users, feel the need to utter their feelings and impressions (both when they are happy and when they are unhappy), are well informed about product and brands and eager to convey this information to others. Individuals who lack these traits will likely not comment about their purchases in the online environment, will not praise or blast products, even if they had a particularly good or bad experience in the market or if they have a special relationship with a brand. People of this kind just don't have the disposition to communicate their experiences and sentiments in the social networks.

Our research also examines the relationships between eWOM propensity and the intention to re-purchase the product. The findings show that positive eWOM intentions are positively associated with the intention to re-purchase, while negative eWOM intentions are negatively related with the same variable. This result is supported by the study of (East, Uncles, Romaniuk, & Lomax, 2016). It is expected that social network members don't recommend products or services they don't use themselves, so they will probably buy those products again in the future. On the other hand, people who complain about products online will be undoubtedly discouraged from repeating the purchase. (East, Uncles, Romaniuk, & Lomax, 2016) also show that positive eWOM has more impact than negative eWOM on the intention to repurchase, which is in concordance with our results (the coefficient of the positive eWOM propensity is bigger in absolute value).

**Limitations and Further Research**

This research suffers from several limitations. First, we have used the convenience sampling method to select our survey respondents. Second, our sample comprises Romanian subjects only, most of them students aged under 26 (because people in this category are intense social media users). These limitations might negatively affect, to some extent, the generalization of our results. Future research could examine other segments of the population in terms of age, education or profession.

To elucidate some contradictory issues and gain more understanding about the foremost predictors of eWOM propensity, further studies could investigate the moderating effect of the customers' personality traits in the relationship between product related factors and eWOM intentions. Antecedent variables like product quality, customer satisfaction, brand attitude or price perception could be considered for these studies.

**Conclusions and Practical Implications**

The goal of our research was to reveal the most important factors that drive social media users to create positive or negative WOM in the online environment. We have considered two types of influencing variables: product related factors (customers' perception regarding the product quality and price, as well as their attitude towards the brand) and personal traits (people's intrinsic characteristics). The major theoretical contribution of this study consists of revealing that personal traits are stronger eWOM propensity predictors than product related variables. In the light of our results, most people will likely not share their positive or negative opinions about products and brands, no matter how good or bad their experience was, no matter how satisfied or dissatisfied they are. The high eWOM intention levels are associated with the presence of some inner characteristics that push people to voice their feelings and thoughts in the social networks. As our research shows, the most important of these traits are social media usage (the degree to which a customer uses social media to stay informed about products and companies) and need to evaluation (the urge to have strong opinions about everything and share them with others). People with high scores for these variables are the most likely to engage in eWOM activities, spreading the word about products and brands. Another significant eWOM antecedent is marketing mavenism, but this variable only influences positive eWOM intentions. Marketing mavens use to recommend valuable products to their social network peers and rarely comment about bad products.

According to our findings, the only product related factor that has a noteworthy impact on eWOM propensity is brand trust. Customers with great confidence in the brand have lower probabilities to write negative reviews about it, even if they have reasons of dissatisfaction sometimes. So brand trust only influences the intention to provide negative eWOM.

These results have important practical implications. To get a large amount of positive eWOM about their products, companies should identify customers and supporters with great likelihood of writing favorable reviews and comments online. In other words, they should look for active social media users, mavens, opinion leaders, people who like to talk about their experiences and express their opinions in a loud voice. These people should be encouraged to spread eWOM, because they are the best candidates to become enthusiastic company ambassadors. If they trust the company and enjoy its products, they will probably share the word about them, build credibility for the brand and bring many new customers for free.

Our research also shows that both positive and negative eWOM intentions are associated with the intention to repurchase. More specifically, customers who praise the product in the social media will probably decide to continue buying it, while those who criticize the product will not repeat the purchase. This outcome reveals another important benefit for the companies that create and leverage positive eWOM. Social media users who share positive information about the product and recommend it to others are motivated to buy it again and become loyal customers.

**Appendix -** Construct items
This appendix is based on the previous scales proposed by: (Wu, Lu, Wu, & Fu, 2012), (Chaudhuri & Holbrook, 2001), (Rapp, Beitelspacher, Grewal, & Hughes, 2013), (Steenkamp & Gielens, 2003), (Jarvis & Petty, 1996), (Alexandrov, Lilly, & Babakus, 2013), (Wolter & Cronin, 2016), (Dodds, Monroe, & Grewal, 1991). Some of these scales are also available at: https://www.jean-pfiffelmann.com/marketing-scales/.

**Product quality**
This product is good quality
This product is of high workmanship
This product is of high durable
This product is of high reliable
This product is of high dependable

**Price perception**
This product is very good value for the money*
At the price shown the product is very economical
The product is considered to be a good buy
The price shown for the product is very acceptable
This product appears to be a bargain*

**Brand trust**
I trust this brand
I rely on this brand
This is an honest brand
This brand is safe

**Social media usage**
I use social media to monitor products and brands
I use social media to follow sales and promotions
I use social media to monitor events*
People use social media to reach me*
My relationship with products and brands is enhanced by social media*
I use social media to communicate with retailers*
I use social media to improve my relationship with retailers*
My relationship with my retail store is enhanced by social media*

**Marketing mavenism**
I like introducing new brands and products to my friends.
I use to talk to friends about the products that I buy.*
My friends and neighbors often come to me for advice.
People often ask me for my opinion about new products.

**Need to evaluate**
I form opinions about everything.*
I prefer to avoid taking extreme opinions.*
It is vey important to me to hold strong opinions.*
I want to know exactly what is good and bad about everything.*
I often prefer to remain neutral about complex issues.*
If something does not affect me, I do not usually determine if it is good or bad.*
I enjoy strongly liking and disliking new things.*
There are many things for which I do not have a preference.
It bothers me to remain neutral.
I like to have strong opinions even when I am not personally involved.
I have many more opinions that the average person.
I would rather have a strong opinion that no opinion at all.
I pay a lot of attention to whether things are good or bad.*
I only form strong opinions when I have to.*
I like to decide that new things are really good or really bad.*
I am pretty much indifferent to many important issues.*

**Positive eWOM intention**
I would post positive things about this brand.
I would recommend this brand to people in my social networks
If a member of my network seeks my advice, I will recommend them to buy this product*

**Negative eWOM intention**
I would complain to the members of my social network about this product
I would discuss with the members of my social network about my frustrations
I would say negative things about the brand in my social networks

**Intention to re-purchase**
The likelihood of re-purchasing this product is: (Very low to Very high).
The probability that I would buy this product again is: (Very low to Very high).
My willingness to buy this product again is: (Very low to Very high).

* this item was removed in the process of scale purification